# Massively parallel and universal approximation of nonlinear functions using diffractive processors

Md Sadman Sakib Rahman[1,2,3]   mssr@ucla.edu
Yuhang Li[1,2,3]                yuhangli@ucla.edu
Xilin Yang[1,2,3]               mikexlyang@ucla.edu
Shiqi Chen[1,2,3]               chensq0120@ucla.edu
Aydogan Ozcan[1,2,3,*]           ozcan@ucla.edu

[1]Electrical and Computer Engineering Department, University of California, Los Angeles, CA, 90095, USA
[2]Bioengineering Department, University of California, Los Angeles, CA, 90095, USA
[3]California NanoSystems Institute (CNSI), University of California, Los Angeles, CA, 90095, USA
[*]Corresponding author: ozcan@ucla.edu

## Abstract

Nonlinear computation is essential for a wide range of information processing tasks, yet implementing nonlinear functions using optical systems remains a challenge due to the weak and power-intensive nature of optical nonlinearities. Overcoming this limitation without relying on nonlinear optical materials could unlock unprecedented opportunities for ultrafast and parallel optical computing systems. Here, we demonstrate that large-scale nonlinear computation can be performed using linear optics through optimized diffractive processors composed of passive phase-only surfaces. In this framework, the input variables of nonlinear functions are encoded into the phase of an optical wavefront—e.g., via a spatial light modulator (SLM)—and transformed by an optimized diffractive structure with spatially varying point-spread functions to yield output intensities that approximate a large set of unique nonlinear functions – all in parallel. We provide proof establishing that this architecture serves as a universal function approximator for an arbitrary set of bandlimited nonlinear functions, also covering multi-variate and complex-valued functions. We also numerically demonstrate the parallel computation of one million distinct nonlinear functions, accurately executed at wavelength-scale spatial density at the output of a diffractive optical processor. Furthermore, we experimentally validated this framework using *in situ* optical learning and approximated 35 unique nonlinear functions in a single shot using a compact setup consisting of an SLM and an image sensor. These results establish diffractive optical processors as a scalable platform for massively parallel universal nonlinear function approximation, paving the way for new capabilities in analog optical computing based on linear materials.

# Introduction

Computing has been pivotal in scientific advances and technological breakthroughs, with digital electronics dominating the field due to its scalability and precision. However, the limits of the scaling predicted by Moore's Law[1] and the evolving nature of computation driven by machine learning and artificial intelligence (AI)[2,3] have spurred interest in alternative approaches. Architectures such as artificial neural networks (ANNs) have become popular for AI tasks, e.g., pattern recognition and decision-making, and are becoming one of the dominant sources of energy consumption globally. Analog optical processing can potentially offer energy efficiency and speed advantages for neuromorphic computing architectures[4–8], especially if the information of interest is in the optical domain, such as the visual information contained within a scene. Diffractive processing of optical information in free space by a series of spatially optimized surfaces has gained traction in recent years for dense and efficient neuromorphic computing [9–12]. Interleaved by free space propagation of optical waves, such structured passive materials locally modulate the amplitude and/or the phase of the incident wavefront that carries the information of interest. This modulation can be realized by varying the spatial features (e.g., the thickness profile) of a passive structured material; alternatively, it can be accomplished by using spatial light modulators (SLMs)[13,14]. A statistically optimum modulation profile for a desired computational task can be obtained through data-driven digital training of a diffractive processor model using deep learning[15] tools such as error backpropagation. After the training stage, these optimum profiles can either be translated into the spatial features of physical surfaces using appropriate 3D fabrication or lithography methods[9,16] or be used to drive SLMs[17].

While diffractive optical processing has been shown to provide universal linear processing[18–22], it, in general, lacks an inexpensive, efficient and scalable form of nonlinear processing capability, which is essential for universal computing[23]. Optoelectronic detection by a photodetector or an image sensor array, following the diffractive processing of analog input information in free-space, can provide only second-order nonlinearity, and this is not cascadable. Nonlinear interaction of light waves can be facilitated using nonlinear optical materials; however, this requires very high optical intensities. For example, the field intensities associated with typical nonlinear optical processes, such as second harmonic generation (>$10^{11}$ W/cm$^2$)[24–26], ultrafast nonlinear absorption (>$10^6$ W/cm$^2$)[27,28], and nonlinear Kerr effect (>$10^4$ W/cm$^2$)[29,30], are many orders of magnitude larger than the ambient light intensity captured by digital cameras (typically below 0.1 W/cm$^2$)[31]. Photochromic[32,33] and photorefractive[34–36] effects may also be used for nonlinear effects, but these processes are typically slow (response time scales on the order of a few[32,34] to tens of seconds[36]). The photorefractive effect based on multiple quantum wells is fast but presents weak nonlinearity[37–39]. Also, photochromic or photorefractive materials generally feature strong absorption (e.g., >90%), leading to substantial losses[40].

There have been recent attempts at performing nonlinear information processing with linear optics[17,41–44]. For example, the reinsertion of the input information at different planes by using SLMs was introduced[17]. This encoding scheme showed success in improved image classification; however, it is not compatible with all-optical information processing using passive surfaces as the input information must be retrieved in each case and digitally pre-processed to re-appear at different planes. Also, it cannot be used for performing an arbitrary linear transformation due to object dependency of the input point-spread functions of the system[41]. In another exciting effort[42], nonlinear optical encoding was demonstrated by recurrent linear scattering from an SLM placed at the boundary of an optical cavity. This scheme performs nonlinear optical encoding of the input information, however, subsequent processing is performed digitally; therefore, it is not an optical function approximator. As another successful example, Wanjura et al.[43] demonstrated encoding the input information in the system parameters—such as the frequency detuning of resonators in a photonic integrated circuit. However, such forms of encoding are relatively complex to implement and fine-tune at a large scale. All this recent progress indicates that implementing nonlinear computation using linear optics is a highly desired capability that could potentially offer transformative paths toward passive, scalable, and high-speed information processing.

In this work, we introduce a framework for massively parallel, universal nonlinear function approximation that relies exclusively on linear optical components, built upon the concept of wavefront encoding followed by spatially optimized diffractive processing. In this approach, the input variables ($a$ or $\vec{a}$) of the nonlinear functions of interest are encoded into the phase of an optical wavefront—for example, using an SLM—and processed by an optimized and fixed diffractive processor (exhibiting spatially varying point-spread functions, PSFs) to yield output intensities that approximate the desired nonlinear functions at the diffraction limit of light. A distinctive advantage of this architecture is its extreme parallelism: each diffraction-limited output pixel of a diffractive processor can be allocated to a unique nonlinear function, enabling the simultaneous computation of tens of thousands to millions of nonlinear functions with a single optical processor. We establish the universality of this diffractive computing paradigm through theoretical analysis, proving that any arbitrary set of bandlimited nonlinear functions, including multi-variate and complex-valued functions, can be approximated via optimized diffractive processing of encoded wavefronts. We numerically demonstrate parallel computation of one million arbitrary nonlinear functions at wavelength-scale spatial density using a single diffractive processor; we also validate this concept experimentally using *in situ* optical learning through a compact optical setup comprising a single phase-only SLM and an image sensor, executing 35 distinct nonlinear functions all in parallel. With contemporary image sensors capable of capturing hundreds of megapixels to gigapixels, this framework offers exceptional parallel processing capability at the diffraction limit of light, potentially allowing simultaneous computation of hundreds of millions of nonlinear functions within a compact diffractive processor. Together, these results establish diffractive optical processors as passive and massively parallel hardware platforms for universal nonlinear computation—ushering in a new class of analog optical systems that harness linear materials for nonlinear computing.

# Results

We schematically depict the nonlinear function implementation framework in **Fig. 1**, illustrating the wavefront encoding and subsequent diffractive processing. As shown in **Fig. 1a**, the diffractive processor is assumed to comprise $N$ optimized (and fixed) phase features/elements distributed over $K$ surfaces. $N_p$ pixels at the input aperture of the diffractive processor are dynamic and are used to encode the input $a$ that represents the argument of the nonlinear functions to be computed, whereas the intensities at the $N_f$ output detectors/pixels represent the optically computed values of the target functions $f_k(a)$ by the diffractive processor, where $k = 1, \cdots, N_f$. The number of pixels available at the input, i.e., $N_p$, sets the bandwidth of the functions that can be computed by the diffractive processor, whereas the number of (diffraction-limited) output pixels $N_f$ determines the theoretical limit of nonlinear function multiplexing; see the '**Theoretical analysis**' section below.

**Figure 1b** further illustrates this concept by showing example phase patterns $\varphi_{in}(p; a)$ for different values of $a$ with $N_p = 9$. Here, the phase encoding is determined by $\varphi_{in}(p; a) = 2\pi(p - 1)a$. For a given value of $a$ and the corresponding input phase pattern, the normalized optical intensities at the $N_f$ output pixels of the optimized diffractive processor follow the respective target functions. In other words, the $k$-th output pixel traces the shape of a unique nonlinear function $f_k(a)$ where $k = 1, \cdots, N_f$; and this nonlinear function approximation is realized simultaneously for all $N_f$ functions (e.g., $N_f = 10^6$) in a single optical inference step—executed in parallel within a few tens of picoseconds through a thin optical volume from the input plane to the output plane. To elucidate the fundamental operating principles and capabilities of this framework, we begin with a theoretical analysis of its forward optical model, detailed next.

## Theoretical analysis

Without loss of generality, we begin by analyzing the forward optical model for 1D nonlinear functions $f(a)$. This analysis also generalizes to higher-dimensional and complex-valued functions, as discussed below. We consider practically relevant functions that are bandlimited, with a Fourier bandwidth of $\Delta_\alpha = \alpha_{max} - \alpha_{min}$, such that $F(\alpha) = 0$ for $\alpha < \alpha_{min}$ or $\alpha \geq \alpha_{max}$, where $F(\alpha) = \int f(a)e^{-j2\pi\alpha a}da$ is the Fourier transform of $f(a)$. Such bandlimited functions can be represented using a finite number $N_\alpha$ of Fourier coefficients $F(\alpha_i)$, where $i = 1, \cdots, N_\alpha$. Our goal is to approximate $N_f$ such functions $f_k(a)$ that are arbitrarily defined in terms of their Fourier coefficients $F_k(\alpha_i)$, i.e.,

$$f_k(a) = \sum_{i=1}^{N_\alpha} F_k(\alpha_i)e^{j2\pi\alpha_i a} \qquad (1)$$

where $k = 1, \cdots, N_f$. All these $N_f$ nonlinear functions will be optically computed by an optimized and fixed diffractive processor at $N_f$ diffraction-limited pixels on the output plane. Even if the target functions are not bandlimited, one can establish a sufficiently accurate approximation to any $f_k$ by using a finite number of Fourier frequencies in **Eq. 1**.

Let $N_p$ denote the number of pixels on the input plane. The optical field at output pixel $k$ is given by $u_{out}(k) = \sum_{p=1}^{N_p} \hat{F}(k;p)u_{in}(p)$, where $u_{in}(p)$ is the optical field phasor at input pixel $p$ and $\hat{F}(k;p)$ is the complex-valued diffractive connection weight between input pixel $p$ and output pixel $k$, determined by the *spatially varying coherent PSFs* [18,19,45] of the diffractive processor. To explicitly express the dependence on the input $a$ of the functions to be computed, we write: $u_{out}(k;a) = \sum_p \hat{F}(k;p)u_{in}(p;a)$. Assuming a phase-only SLM at the input, which sets the phase at pixel $p$ as $\varphi_{in}(p;a) = 2\pi\alpha_p a$, we get $u_{in}(p;a) = e^{j2\pi\alpha_p a}$ and

$$u_{out}(k;a) = \sum_p \hat{F}(k;p)e^{j2\pi\alpha_p a} \qquad (2)$$

Since diffractive optical processors can be optimized to synthesize arbitrarily defined $\hat{F}(k;p)$[19], by setting $\hat{F}(k;p) \approx F_k(\alpha_p)$ and defining the optically computed functions as $\hat{f}_k(a) = u_{out}(k;a)$, we have

$$\hat{f}_k(a) \approx \sum_p F_k(\alpha_p)e^{j2\pi\alpha_p a} = f_k(a) \qquad (3)$$

Hence, with a sufficiently large $N_p$, arbitrarily defined bandlimited nonlinear functions $f_k(a)$ can be optically computed, all in parallel, at the output of an optimized and fixed diffractive optical processor. This framework can also be extended to multi-variate functions $f_k(a_1, \cdots, a_D)$ by assigning each input pixel to a unique $D$-dimensional Fourier frequency $(\alpha_1, \cdots, \alpha_D)$.

These analyses indicate that diffractive optical processors can be used to approximate complex-valued bandlimited nonlinear functions if the phase of the output optical wave can be measured; however, without loss of generality, for the rest of the manuscript, we only consider real-valued nonlinear functions, which can be measured with detectors that are sensitive only to optical intensity. In an alternative strategy, part of $N_f$ output pixels can be assigned to the real and imaginary parts of complex-valued nonlinear functions to implement them all in parallel using intensity-based output detectors.

This analysis above should **not** be confused with the Fourier transforming property of free-space diffraction or lenses since such **spatially invariant transformations** apply to the spatial frequencies of the field over an input aperture to perform the spatial Fourier transform operation at the output. In our analysis described above, there are $N_f$ (e.g., $N_f = 10^6$) unique nonlinear functions that are arbitrarily defined, and their Fourier frequencies do **not** represent the spatial frequencies of the input fields carried by propagating waves. Unlike a spatial Fourier transform operator, the optimized diffractive processor, in our case, has a **spatially varying complex-valued PSF set defined by** $\widehat{F}(k; p)$. In the case of standard free-space diffraction or lens-based Fourier processors, the system boils down to a spatially invariant convolution operator, and it cannot execute in parallel a set of unique nonlinear functions between the input and output, as we demonstrate in this work.

## Numerical analysis

For the numerical simulation results that follow, without loss of generality, we assume an operating wavelength of $\lambda$ = 550 nm. However, the analyses and conclusions presented here are not wavelength-specific and are applicable across the electromagnetic spectrum, provided that all physical dimensions—such as axial spacing and lateral pitch—are scaled proportionally with the illumination wavelength. For these simulations, we assume $N_p = 9$, where the $N_p$ input pixels are arranged contiguously in a square grid at a lateral pitch $\delta$ of 300 nm (~$0.55\lambda$), as illustrated in **Fig. 1b**. At the output aperture, $N_f$ output pixels (each having a lateral width of $\delta$, i.e., ~$0.55\lambda$) are similarly arranged in a square grid. To suppress cross-talk between neighboring outputs, an inter-pixel spacing of ~$0.5\lambda$ is included—see **Fig. 1a**. The width of each optimizable phase-only diffractive feature/element comprising the diffractive surfaces is assumed to be $\delta$. The axial distance between consecutive planes, whether they accommodate the diffractive surfaces or the input/output pixels, is set as $W\sqrt{\left(\frac{2\delta}{\lambda}\right)^2 - 1}$ where $W = \sqrt{\frac{N}{K}}\delta$ is the total width of a diffractive surface. All designs use the same number of input pixels, $N_p = 9$, which also sets the bandwidth of the target nonlinear functions. The input variable $a$ of the functions is encoded using the same phase encoding scheme across all the designs, i.e., $\varphi_{in}(p; a) = 2\pi(p - 1)a$ where $p = 1, \cdots, N_p$.

To evaluate the scalability and approximation accuracy of our diffractive processing framework, we designed five separate optical processors for increasing numbers of target unique nonlinear functions, $N_f$, ranging from $10^2$ to $10^6$. For each design, the number of trainable diffractive features was set to $N \approx 1.25 \times 2N_pN_f$, distributed evenly across $K = 2$ surfaces. **Figure 2a** presents the distribution of function approximation errors (see **Eq. 16** in the **Methods** section for its definition) for all five optical processor designs. As the number of output functions $N_f$ increases, both the spread and the maximum of the error distribution increase, reflecting the growing complexity of the task. Despite this trend, the errors, including for $N_f = 10^6$ remain remarkably low across all designs, underscoring the inherent scalability of the diffractive processor architecture for massively parallel nonlinear function execution.

To further illustrate the performance, **Figures 2b and 2c** display specific examples of optically computed functions from each of the three larger-scale designs: $N_f = 10^4$, $N_f \approx 10^5$ and $N_f = 10^6$. In each case, we show the target nonlinear function (green curve) and the corresponding diffractive approximation (red curve). **Figure 2b** highlights the functions with the lowest approximation error from each design, demonstrating perfect agreement between the target and the diffractive output. **Figure 2c** shows the functions with the largest approximation error from the same three designs. Notably, even in these worst-case function examples, the differences between the target and the approximation are indistinguishable, further confirming the robustness of the proposed method—even for implementing one million parallel nonlinear functions.

In **Fig. 3** and **Supplementary Figs. S1-S3**, we present additional details about some of these diffractive nonlinear function approximators. In **Fig. 3a**, the optimized and fixed surface phase profiles of these diffractive processors are shown. We also show the spatial distributions of the function approximation errors at the output planes of these processors in **Fig. 3b**. At each output pixel location, the approximation error of the corresponding function is shown following the accompanying colormap. Note that the spatial distributions of the function approximation errors reveal relatively larger errors near the center of the output planes for the processors with $K = 2$ surfaces, which can be attributed to the effect of ballistic photons—light waves that propagate through the diffractive volume with minimal modulation. These unperturbed components hinder the performance at the central output pixels, particularly when the processor lacks sufficient structural depth. For the designs in **Fig. 3b** with $K = 2$ surfaces, the diffractive processor faces challenges to adequately redirect the ballistic photons, hence the relatively larger error at the central pixels; however, it is important to note that even the highest error made for implementing $N_f = 10^6$ nonlinear functions is negligible, as indicated in **Fig. 2c**. **Supplementary Figs. S1-S3** show additional 72 representative (randomly selected) examples from the $N_f = 10^6$ nonlinear functions performed by the $K = 2$ diffractive processor.

The overall performance of nonlinear function approximation can be further improved by using a deeper diffractive architecture with larger $K$. In **Fig. 3**, we compare the spatial error distributions of two diffractive nonlinear function approximators with $N_f \approx 10^5$ (two right panels in **Fig. 3b**); these processors have the same number $N$ of optimized/fixed features, distributed over $K = 2$ and $K = 4$ surfaces, respectively. Despite having an equal number of trainable parameters, the shallower design ($K = 2$) exhibits relatively larger approximation errors near the center of the output plane. In contrast, the $K = 4$ design benefits from structural depth, allowing it to reroute and redistribute the ballistic photons, using them more effectively to execute $N_f$ nonlinear functions in parallel. As a result, the approximation errors become much smaller and spatially uniform, as shown in **Fig. 3b**; also see **Fig. 4,** which supports the same conclusions.

To further investigate how the number of diffractive surfaces and the total number of trainable features affect the nonlinear function approximation accuracy and energy efficiency of the nonlinear diffractive processing framework, we simulated multiple architectures with $K = 2$ and $K = 4$ diffractive layers. For each configuration, we varied the total number of optimizable phase-only features and evaluated the resulting performance on a fixed set of $N_f = 10^4$ target nonlinear functions. **Figure 4a** presents the mean and maximum function approximation errors as a function of the total number of diffractive features. As expected, both error metrics decrease with increasing diffractive feature count, reflecting the enhanced expressivity of higher-capacity diffractive optical networks. Importantly, designs with $K = 4$ consistently outperform those with $K = 2$. This structural depth-induced performance gain demonstrates that increasing the number of diffractive layers enables more efficient use of the optical degrees of freedom and facilitates more accurate approximation of large sets of unique nonlinear functions.

Furthermore, **Figure 4b** shows the average output diffraction efficiency (see **Eq. 17** in the **Methods** section for its definition) of the same designs as in **Fig. 4a**. As the number of diffractive features increases, the diffraction efficiency in general improves for both values of $K$, with $K = 4$ consistently yielding higher diffraction efficiencies. This result highlights another important advantage of deeper diffractive architectures: in addition to improved accuracy, they also enable more efficient light throughput, making them better suited for practical implementations where optical power is a critical consideration.

In fact, the function approximation performance of diffractive processors can be traded off for improved output diffraction efficiency by engineering the training loss function. To demonstrate this, we introduced a modified loss function that incorporates a penalty term for low diffraction efficiency. Specifically, the new training loss is defined as $L = L_{PSF} + L_{DE}$, where $L_{DE} = \max(0, DE_{trgt} - DE_{PSF})$. Here, $DE_{PSF}$ denotes the average diffraction efficiency and $DE_{trgt}$ is a user-defined threshold; see the **Methods** section for details. The additional penalty term $L_{DE}$ is only activated during training when the average diffraction efficiency falls below the target value specified by $DE_{trgt}$, thereby encouraging energy-efficient designs without fundamentally altering the optimization framework. **Figure 5a** shows the mean and maximum function approximation errors for four diffractive processor designs, each trained with a different value of $DE_{trgt}$. As expected, higher values of $DE_{trgt}$ lead to higher output diffraction efficiencies $DE$, as desired, but also result in relatively increased approximation errors. This highlights the trade-off between energy efficiency and nonlinear function approximation performance, with stronger constraints on diffraction efficiency slightly degrading the accuracy. To assess the worst-case behavior under the most power-efficient configuration, **Figure 5b** presents the target function and its diffractive optical approximation for the function associated with the maximum error in the highest-efficiency design marked 'I' in **Fig. 5a**. Despite the relatively larger error, the approximation remains accurate across most of the input domain. The perceptible deviations only occur near the domain boundaries (e.g., $a = -0.5$ and $a = 0.5$).

## Experimental results

In our experimental validation of nonlinear function execution using diffractive optical processors, we used a single SLM and an image sensor-array, where a subset of the SLM pixels is allocated for input encoding, while the remaining pixels function as trainable diffractive phase elements that are optimized and fixed. To mitigate the impact of alignment errors and other physical imperfections that typically arise when transferring digitally optimized designs to physical hardware, we adopted an *in situ* learning strategy for experimental training and validation; see the **Methods** section. **Figure 6** presents the experimental results of *in situ* training for a single diffractive layer designed to perform 35 distinct nonlinear functions. As illustrated in **Fig. 6a**, a collimated laser beam (520 nm) illuminates a phase SLM, which simultaneously serves as both the input encoder and the trainable diffractive layer. The modulated wavefront then propagates through free space and forms interference patterns at designated detection regions, each assigned to a unique function, which are captured by a camera. To enable function-specific learning, the central 96×96-pixel region of the SLM was allocated for encoding the nonlinear function input $(a)$, while the surrounding pixels within a 1080×1920-pixel region were the learnable diffractive features trained *in situ* (see the **Methods** section for details). We also predefined 70 detection regions—35 signal regions ($S_1$–$S_{35}$) and 35 corresponding reference regions ($R_1$–$R_{35}$), as shown in **Fig. 6b**—to mitigate variations/fluctuations in the illumination intensity.

**Figure 6c** shows an example predefined input phase pattern corresponding to a scalar input value of $a = 0.64$, encoded within the central 96×96 region of the SLM. The phase values in this region encode the input $a$ of the nonlinear functions and are not optimized during training – in other words, they are not learnable parameters and represent the input values for the nonlinear functions of interest. The surrounding region presents the learned phase profile of the diffractive layer, which was optimized *in situ* and remained fixed after convergence. The evolution of the training process is illustrated in **Fig. 6d**, demonstrating stable convergence throughout the *in situ* optimization. We adopted a progressive training strategy, starting with 5 nonlinear functions and gradually increasing the number of functions by 5 at a time up to $N_f = 35$. Eventually, all 35 functions were jointly optimized *in situ* to achieve the final performance (see the **Methods** section for details). **Figure 7** summarizes the system's performance across all 35 target functions. Each subplot displays the experimentally measured outputs (red dots) overlaid with the target nonlinear functions (green dots). Across all these unique nonlinear functions, the measured outputs show strong agreement with the desired responses, achieving a mean absolute error (MAE) ~0.06. We also performed model-based *in silico* training for the same hardware setup as was used for the *in situ* training, targeting the same set of 35 nonlinear functions. The numerical results corresponding to this ideal (model-based) training are shown with blue dots in **Fig. 7**. While the *in situ* (experimental) training results exhibit occasional deviations due to experimental nonidealities such as vibrations, misalignments, and measurement noise, the numerical results based on the *in silico* training are indistinguishable from the target functions, as desired. In addition to the 35-function diffractive model, we also trained a separate model *in situ* to experimentally perform 16 nonlinear functions simultaneously. For this task, all 16 functions were trained collectively from the beginning, as opposed to the progressive *in situ* learning case highlighted in **Fig. 6d**. The results, shown in **Supplementary Figs. S4**, further support the robustness and flexibility of our training framework under different *in situ* learning strategies.

In our experiments, we limited the *in situ* learning to $N_f = 35$ due to the slow convergence of *in situ* training using the architecture of **Fig. 6a**, which is hindered by experimental nonidealities such as mechanical vibrations and measurement noise. We also conducted model-based *in silico* training of diffractive phase elements for the same single SLM architecture targeting $N_f = 100$ and $N_f = \sim 10^5$; the corresponding results are presented in **Supplementary Figs. S5-S9**. Owing to its limited structural depth, the experimental setup in **Fig. 6a** is less expressive than deeper diffractive architectures with e.g., $K = 2-4$ layers reported in our earlier analyses; however, the near-perfect approximation of $N_f = 100$ and $N_f = \sim 10^5$ nonlinear functions, achieved using a modest number of SLM pixels, demonstrates the utility of this simplified experimental architecture for performing an arbitrary set of nonlinear functions. For the case of $N_f \approx 10^5$ functions approximated using this shallow diffractive architecture (see **Supplementary Figs. S7–S9**), we employed the arrangement shown in **Supplementary Fig. S7a**, where the unit showing a pair of reference and signal detectors (corresponding to a unique function) are repeated $N_f$-times along a square grid; this architecture helped us accommodate a larger number of detectors on the sensor plane. At the SLM plane, the tile containing the input encoding region with surrounding optimizable diffractive elements is repeated similarly, resulting in equal fields of view at the SLM and sensor planes. **Supplementary Figure S7b** shows the optimized (static) diffractive layer for the parallel implementation of $N_f \approx 10^5$ nonlinear functions, where the white regions denote the SLM pixels allocated for dynamic input encoding ($a$). **Supplementary Figures S8 and S9** show 64 representative (randomly selected) examples from the $N_f \approx 10^5$ nonlinear functions performed by this optimized shallow diffractive processor, demonstrating excellent agreement between the target functions and their corresponding diffractive approximations, emphasizing the scalability of this shallow architecture.

Overall, these results validate the effectiveness of our *in situ* training framework and demonstrate the feasibility of implementing multiple nonlinear functions using a compact, programmable optical layer, enabled by a single SLM.

# Discussions

We introduced a diffractive optical computing framework that enables universal nonlinear function approximation using linear diffractive materials. By combining phase-based wavefront encoding with optimized diffractive transformations, our approach achieves extreme levels of parallelism—scaling from hundreds to millions of unique nonlinear functions—all computed at the speed of light propagation through a thin optical volume with a parallel inference time on the order of tens of picoseconds. These results open new opportunities for massively parallel nonlinear optical computing, laying the groundwork for future advancements in large-scale optical information processing.

In our approach, the input variable $a$ of the target functions is encoded into the phase of the optical wavefront using a single SLM, where each of the $N_p$ input pixels realizes a phase modulation $\varphi_{in}(p;a) = 2\pi\alpha_p a$. This generates a spatially structured (encoded) input wavefront comprising $e^{j2\pi\alpha_p a}$. The optimized diffractive processor then performs a linear transformation, i.e., a weighted sum over these components, with learned complex-valued weights corresponding to the coefficients $F_k(\alpha_p)$ of the target nonlinear functions. This optical strategy—realizing a Fourier basis expansion through wavefront modulation, followed by complex-valued linear summation via diffraction—bears conceptual similarity to the "kernel trick" in machine learning[46], where nonlinear mappings are achieved through linear operations in a high-dimensional feature space. In contrast to conventional kernels, however, our system executes this transformation physically, relying solely on linear optical interactions without electronic or optical nonlinearity.

The complexity of the nonlinear functions that can be implemented within our diffractive framework is directly governed by the number of pixels used to encode the function input. A larger number of input pixels $N_p$ enables the representation of a larger number of Fourier bases, thereby supporting the approximation of functions with higher bandwidths exhibiting rapid fluctuations. This framework also generalizes to approximate multi-variate nonlinear functions $f_k(a_1, \cdots, a_D)$ by assigning each input pixel $p$ a unique $D$-dimensional Fourier frequency, i.e., $\varphi_{in}(p;\vec{a}) = 2\pi(\vec{\alpha}_p \cdot \vec{a})$, where $p = 1, \cdots, N_p$, $\vec{a}$ refers to the $D$-dimensional input variable and $\vec{\alpha}_i$ refers to the frequencies within the bandwidth of the target multi-variate nonlinear functions, $f_k(\vec{a})$, $k = 1, \cdots, N_f$.

In our implementation, we normalize all target function values to lie within the [0, 1] range, and apply a similar normalization to the optical intensities representing the function values. This rescaling can be adapted to extend the framework to signed functions that can assume negative values (e.g., [-1, 1] ). For signed functions, alternative strategies might also be employed, such as using a differential configuration with two pixels per function (positive and negative), or referencing output intensities to a global baseline pixel, enabling the representation of signed outputs using strictly positive, i.e., intensity-based optical signals. Similarly, an arbitrary set of complex-valued nonlinear functions can also be implemented all in parallel using intensity-based output detectors by assigning some of these output detectors to the real and imaginary parts of different complex-valued nonlinear functions of interest.

In general, a diffractive optical processor design with an insufficient number of diffractive features (i.e., $N < 2N_p N_f$ ) will trade off nonlinear function approximation accuracy in favor of design simplicity, whereas with a sufficiently large number of diffractive features that are optimized (i.e., $N \geq 2N_p N_f$ ), the approximation error becomes negligible (see **Fig. 4**). Stated differently, the multiplication of the space bandwidth product at the input aperture ($\propto N_p$) with the space bandwidth product at the output aperture ($\propto N_f$) dictates the complexity of the problem at hand, and in general, a larger number of functions that are each higher dimensional (i.e., $f_k(a_1, \cdots, a_D)$) would require a larger number of diffractive degrees of freedom to be optimized. The factor of two in the condition $N \geq 2N_p N_f$ is due to the fact that each diffractive feature within the optical processor to be optimized is a phase-only variable. For a Gedanken experiment, if one could optimize the complex transmission coefficients of the diffractive layers with independent control over both the phase and amplitude of each diffractive feature, one would need 2-fold less diffractive features, i.e., $N \geq N_p N_f$ would be the sufficient condition for convergence.

The number of nonlinear functions that can be computed by our diffractive framework is fundamentally determined by the number of diffraction-limited output pixels ($N_f$), which sets an upper bound for function multiplexing at a single illumination wavelength. By using multiple illumination wavelengths[45], one can further improve the number of unique nonlinear functions that can be implemented through the same diffractive processor if sufficiently large trainable degrees of freedom are available, i.e., $N \geq 2N_w N_p N_f$, where $N_w$ is the number of wavelengths. In this multi-wavelength architecture, without the need for material dispersion engineering, $N_w N_f$ unique nonlinear functions can be implemented in parallel, quantified through the spectra of $N_f$ output detectors; in alternative implementations, vertically stacked photodetector arrays[47] can also be used to read out $N_w N_f$ unique nonlinear functions in parallel. With modern image sensors offering hundreds of megapixels to gigapixels[48-50], this framework could support extremely high degrees of parallelism at the diffraction limit of light—potentially enabling the simultaneous computation of hundreds of millions of functions within a thin diffractive processor. Achieving such massively parallel computation of nonlinear functions requires scaling the physical size and resolution of diffractive surfaces accordingly, which introduces challenges in fabrication fidelity and alignment over larger areas. Nonetheless, pushing the boundaries of fabrication scalability—through emerging techniques in large-area nanofabrication and high-resolution additive manufacturing[51-54]—can unlock unprecedented levels of parallelism in future diffractive optical computing systems.

# Methods

## Evaluation metrics

The function approximation error for the $k$-th nonlinear function was quantified using a normalized root-mean-square error metric defined as:

$$\varepsilon_k = \sqrt{\int \left(f_k(a) - \hat{f}_k(a)\right)^2 da}$$

for $k = 1, \cdots, N_f$.

The average diffraction efficiency $DE$ of the diffractive processor was computed as:

$$DE = \frac{\int DE(a) da}{\int da}$$

where the diffraction efficiency $DE(a)$ for a given value of the functions' input $a$ is defined by:

$$DE(a) = \frac{\sum_{k=1}^{N_f} P_k(a)}{\sum_{p=1}^{N_p} |u_{in}(p;a)|^2}$$

## Training

The diffractive processor models were implemented and trained using PyTorch (version 1.10) with Python 3.9. Training time depended on several factors, including the number of functions $N_f$ and available GPU resources. For example, the $N_f = 10^6$ model was trained for ~1.6K epochs using a single NVIDIA A100 GPU. Each epoch—defined as 1000 iterations of gradient descent on the PSF-based loss (see the Supplementary Information)—took ~20 minutes. Diffractive models with smaller $N_f$ were trained on other machines equipped with GPUs, such as the NVIDIA GeForce RTX 2080 Ti and RTX 4090.

# Supplementary Information

Supplementary Information file includes:

- Supplementary Figures S1-S9
- Supplementary Methods on "*Forward model of diffractive processors for nonlinear function approximation*"
- Supplementary Methods on "*Spatially varying coherent PSF-based diffractive processor optimization for nonlinear function implementation*"
- Supplementary Methods on "*In situ learning for experimental demonstration of nonlinear function implementation*"

# References

1. Shalf, J. The future of computing beyond Moore's Law. *Philos. Trans. R. Soc. Math. Phys. Eng. Sci.* **378**, 20190061 (2020).

2. Fahlman, S. E., Hinton, G. E. & Sejnowski, T. J. Massively parallel architectures for AI: netl, thistle, and boltzmann machines. in *Proceedings of the Third AAAI Conference on Artificial Intelligence* 109–113 (AAAI Press, Washington, D.C., 1983).

3. Stiefel, K. M. & Coggan, J. S. The energy challenges of artificial superintelligence. *Front. Artif. Intell.* **6**, 1240653 (2023).

4. Marković, D., Mizrahi, A., Querlioz, D. & Grollier, J. Physics for neuromorphic computing. *Nat. Rev. Phys.* **2**, 499–510 (2020).

5. The Future of Deep Learning Is Photonic. *IEEE Spectrum* https://spectrum.ieee.org/the-future-of-deep-learning-is-photonic (2021).

6. Li, C., Zhang, X., Li, J., Fang, T. & Dong, X. The challenges of modern computing and new opportunities for optics. *PhotoniX* **2**, 20 (2021).

7. Li, R. *et al.* Photonics for Neuromorphic Computing: Fundamentals, Devices, and Opportunities. *Adv. Mater.* **n/a**, 2312825.

8. Shastri, B. J. *et al.* Photonics for artificial intelligence and neuromorphic computing. *Nat. Photonics* **15**, 102–114 (2021).

9. Lin, X. *et al.* All-optical machine learning using diffractive deep neural networks. *Science* **361**, 1004–1008 (2018).

10. Bai, B. *et al.* To image, or not to image: class-specific diffractive cameras with all-optical erasure of undesired objects. *eLight* **2**, 14 (2022).

11. Luo, Y. *et al.* Computational imaging without a computer: seeing through random diffusers at the speed of light. *eLight* **2**, 4 (2022).

12. Hu, J. *et al*. Diffractive optical computing in free space. *Nat. Commun.* **15**, 1525 (2024).

13. Bartlett, T. A. *et al*. Recent advances in the development of the Texas Instruments phase-only microelectromechanical systems (MEMS) spatial light modulator. in *Emerging Digital Micromirror Device Based Systems and Applications XIII* vol. 11698 103–116 (SPIE, 2021).

14. Yang, Y., Forbes, A. & Cao, L. A review of liquid crystal spatial light modulators: devices and applications. *Opto-Electron. Sci.* **2**, 230026–29 (2023).

15. LeCun, Y., Bengio, Y. & Hinton, G. Deep learning. *Nature* **521**, 436–444 (2015).

16. Bai, B. *et al*. Data-Class-Specific All-Optical Transformations and Encryption. *Adv. Mater.* **35**, 2212091 (2023).

17. Yildirim, M., Dinc, N. U., Oguz, I., Psaltis, D. & Moser, C. Nonlinear processing with linear optics. *Nat. Photonics* **18**, 1076–1082 (2024).

18. Kulce, O., Mengu, D., Rivenson, Y. & Ozcan, A. All-optical information-processing capacity of diffractive surfaces. *Light Sci. Appl.* **10**, 25 (2021).

19. Kulce, O., Mengu, D., Rivenson, Y. & Ozcan, A. All-optical synthesis of an arbitrary linear transformation using diffractive surfaces. *Light Sci. Appl.* **10**, 196 (2021).

20. Rahman, M. S. S., Yang, X., Li, J., Bai, B. & Ozcan, A. Universal linear intensity transformations using spatially incoherent diffractive processors. *Light Sci. Appl.* **12**, 195 (2023).

21. Yang, X., Rahman, M. S. S., Bai, B., Li, J. & Ozcan, A. Complex-valued universal linear transformations and image encryption using spatially incoherent diffractive networks. *Adv. Photonics Nexus* **3**, 016010 (2024).

22. Rahman, M. S. S. & Ozcan, A. Universal point spread function engineering for 3D optical information processing. *Light Sci. Appl.* **14**, 212 (2025).

23. Hornik, K., Stinchcombe, M. & White, H. Multilayer feedforward networks are universal approximators. *Neural Netw.* **2**, 359–366 (1989).

24. Aparajit, C. *et al.* Efficient second-harmonic generation of a high-energy, femtosecond laser pulse in a lithium triborate crystal. *Opt. Lett.* **46**, 3540–3543 (2021).

25. Wang, Y. *et al.* Direct electrical modulation of second-order optical susceptibility via phase transitions. *Nat. Electron.* **4**, 725–730 (2021).

26. Yi, F. *et al.* Optomechanical Enhancement of Doubly Resonant 2D Optical Nonlinearity. *Nano Lett.* **16**, 1631–1636 (2016).

27. Long, H. *et al.* Tuning nonlinear optical absorption properties of WS2 nanosheets. *Nanoscale* **7**, 17771–17777 (2015).

28. Sun, Z. *et al.* Graphene Mode-Locked Ultrafast Laser. *ACS Nano* **4**, 803–810 (2010).

29. Klopfer, E., Lawrence, M., Barton, D. R. I., Dixon, J. & Dionne, J. A. Dynamic Focusing with High-Quality-Factor Metalenses. *Nano Lett.* **20**, 5127–5132 (2020).

30. Chen, X. *et al.* Optical nonlinearity and non-reciprocal transmission of graphene integrated metasurface. *Carbon* **173**, 126–134 (2021).

31. Vatteroni, M., Covi, D. & Sartori, A. A linear-logarithmic CMOS pixel for high dynamic range behavior with fixed-pattern-noise correction and tunable responsivity. in *2008 IEEE SENSORS* 930–933 (2008). doi:10.1109/ICSENS.2008.4716593.

32. Hirata, S., Totani, K., Yamashita, T., Adachi, C. & Vacha, M. Large reverse saturable absorption under weak continuous incoherent light. *Nat. Mater.* **13**, 938–946 (2014).

33. Kobayashi, Y. & Abe, J. Recent advances in low-power-threshold nonlinear photochromic materials. *Chem. Soc. Rev.* **51**, 2397–2415 (2022).

34. Ducharme, S. & Feinberg, J. Altering the photorefractive properties of BaTiO$_3$ by reduction and oxidation at 650°C. *JOSA B* **3**, 283–292 (1986).

35. Xue, L. *et al.* The Photorefractive Response of Zn and Mo Codoped LiNbO3 in the Visible Region. *Crystals* **9**, 228 (2019).

36. Usui, K., Matsumoto, K., Katayama, E., Akamatsu, N. & Shishido, A. A Deformable Low-Threshold Optical Limiter with Oligothiophene-Doped Liquid Crystals. *ACS Appl. Mater. Interfaces* **13**, 23049–23056 (2021).

37. Partovi, A. *et al.* Cr-doped GaAs/AlGaAs semi-insulating multiple quantum well photorefractive devices. *Appl. Phys. Lett.* **62**, 464–466 (1993).

38. Canoglu, E. *et al.* Carrier transport in a photorefractive multiple quantum well device. *Appl. Phys. Lett.* **69**, 316–318 (1996).

39. Dongol, A., Thompson, J., Schmitzer, H., Tierney, D. & Wagner, H. P. Real-time contrast-enhanced holographic imaging using phase coherent photorefractive quantum wells. *Opt. Express* **23**, 12795–12807 (2015).

40. Khoo, I. C., Wood, M. V., Shih, M. Y. & Chen, P. H. Extremely nonlinear photosensitive liquid crystals for image sensing and sensor protection. *Opt. Express* **4**, 432–442 (1999).

41. Li, Y., Li, J. & Ozcan, A. Nonlinear encoding in diffractive information processing using linear optical materials. *Light Sci. Appl.* **13**, 173 (2024).

42. Xia, F. *et al.* Nonlinear optical encoding enabled by recurrent linear scattering. *Nat. Photonics* **18**, 1067–1075 (2024).

43. Wanjura, C. C. & Marquardt, F. Fully nonlinear neuromorphic computing with linear wave scattering. *Nat. Phys.* **20**, 1434–1440 (2024).

44. Zhang, D. *et al.* Broadband nonlinear modulation of incoherent light using a transparent optoelectronic neuron array. *Nat. Commun.* **15**, 2433 (2024).

45. Li, J. *et al.* Massively parallel universal linear transformations using a wavelength-multiplexed diffractive optical network. *Adv. Photonics* **5**, 016003 (2023).

46. Boser, B. E., Guyon, I. M. & Vapnik, V. N. A training algorithm for optimal margin classifiers. in *Proceedings of the fifth annual workshop on Computational learning theory* 144–152 (Association for Computing Machinery, New York, NY, USA, 1992). doi:10.1145/130385.130401.

47. Foveon X3 sensor. *Wikipedia* (2024).

48. Gray, J. The Biggest Digital Camera Ever Is Ready to Solve Cosmic Mysteries 3,200 Megapixels At a Time. *PetaPixel* https://petapixel.com/2024/04/03/the-biggest-camera-ever-is-ready-to-solve-cosmic-mysteries-3200-megapixels-at-a-time/ (2024).

49. Canon develops CMOS sensor with 410 megapixels, the largest number of pixels ever achieved in a 35 mm full-frame sensor. *Canon Global* https://global.canon/en/news/2025/20250122.html.

50. Fujifilm GFX100 II. *Wikipedia* (2025).

51. Cooper, K. Scalable Nanomanufacturing—A Review. *Micromachines* **8**, 20 (2017).

52. Saha, S. K. *et al.* Scalable submicrometer additive manufacturing. *Science* **366**, 105–109 (2019).

53. Barad, H.-N., Kwon, H., Alarcón-Correa, M. & Fischer, P. Large Area Patterning of Nanoparticles and Nanostructures: Current Status and Future Prospects. *ACS Nano* **15**, 5861–5875 (2021).

54. Shen, C.-Y. *et al.* Broadband Unidirectional Visible Imaging Using Wafer-Scale Nano-Fabrication of Multi-Layer Diffractive Optical Processors. Preprint at https://doi.org/10.48550/arXiv.2412.11374 (2024).

55. Born, M., Wolf, E. & Bhatia, A. B. *Principles of Optics: Electromagnetic Theory of Propagation, Interference and Diffraction of Light*. (Cambridge University Press, 1999).

56. Goodman, J. W. *Introduction to Fourier Optics*. (Roberts and Company Publishers, 2005).

57. Lin, X. *et al.* All-optical machine learning using diffractive deep neural networks. *Science* **361**, 1004–1008 (2018).

58. Mengu, D. *et al*. Misalignment resilient diffractive optical networks. *Nanophotonics* **9**, 4207–4219 (2020).

59. Sutton, R. S., McAllester, D., Singh, S. & Mansour, Y. Policy Gradient Methods for Reinforcement Learning with Function Approximation. in *Advances in Neural Information Processing Systems* vol. 12 (MIT Press, 1999).

60. Zhao, G., Shu, X. & Zhou, R. High-performance real-world optical computing trained by in situ model-free optimization. Preprint at http://arxiv.org/abs/2307.11957 (2023).

# Figures

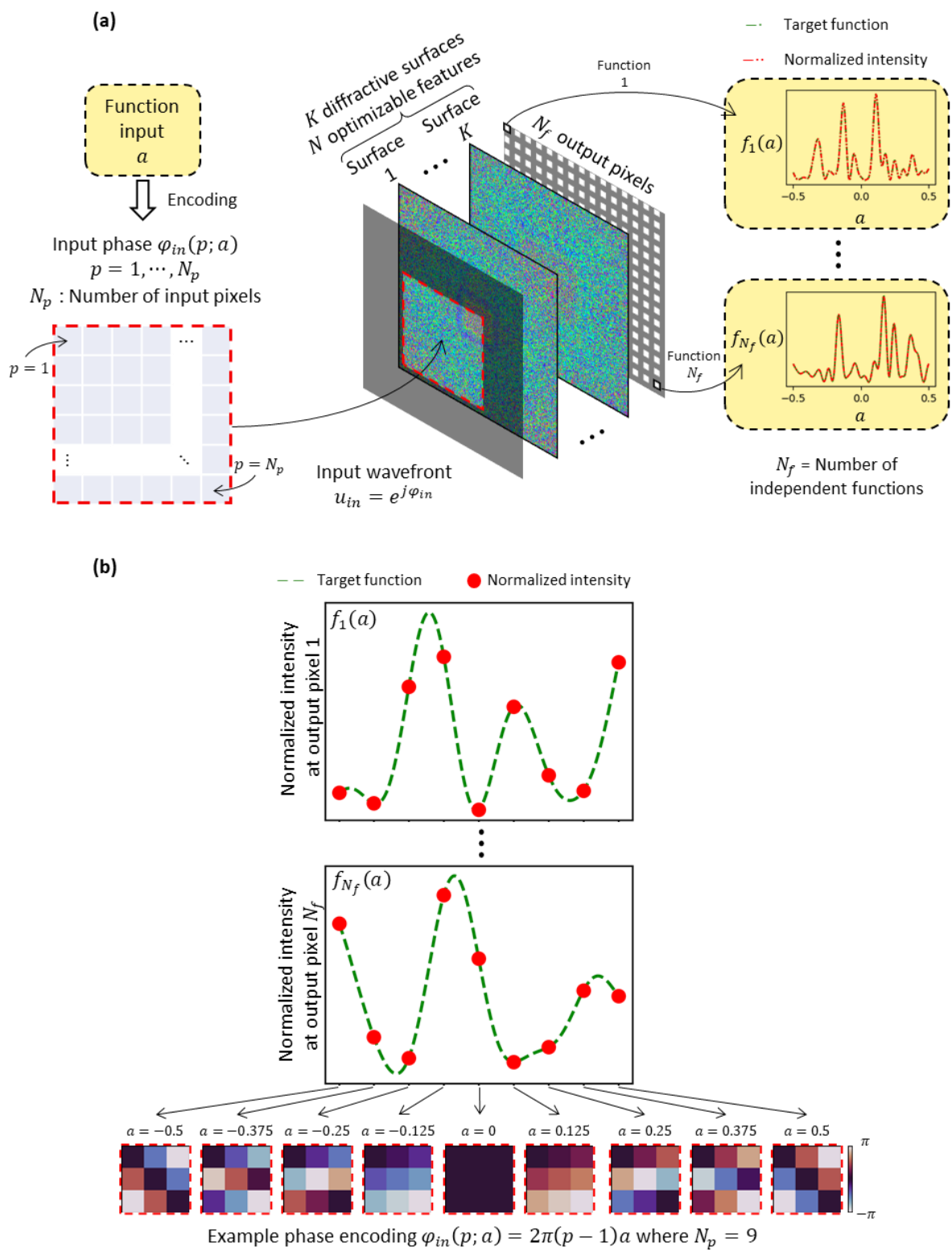

**Fig. 1: Massively parallel diffractive optical computation of nonlinear functions.** (a) The input

variable $a$ of $N_f$ unique nonlinear functions $f_k(a)$ for $k = 1, \cdots, N_f$ is encoded into the phase of the optical wavefront across $N_p$ input pixels. The optical intensities at $N_f$ output pixels of the optimized diffractive processor approximate the output values of these $N_f$ functions. (b) As the phase profile of the input wavefront varies according to $\varphi_{in}(p; a) = 2\pi(p - 1)a$, the normalized optical intensities at the $N_f$ output pixels vary following the target nonlinear functions $f_k(a)$ – performed all in parallel.

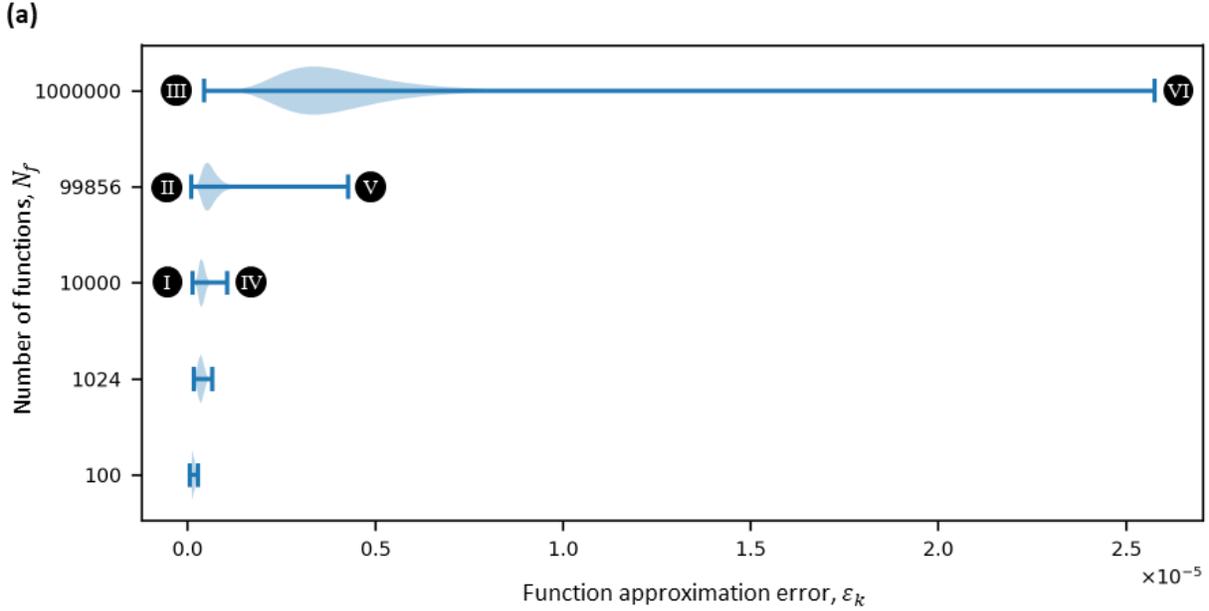

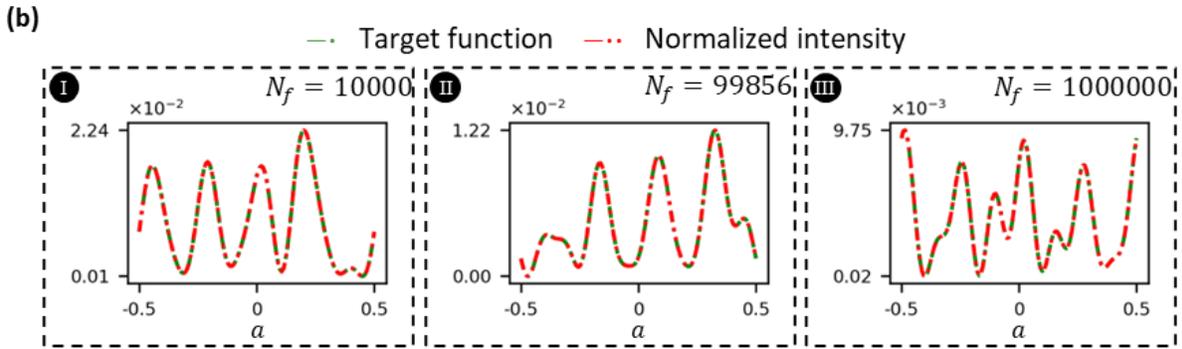

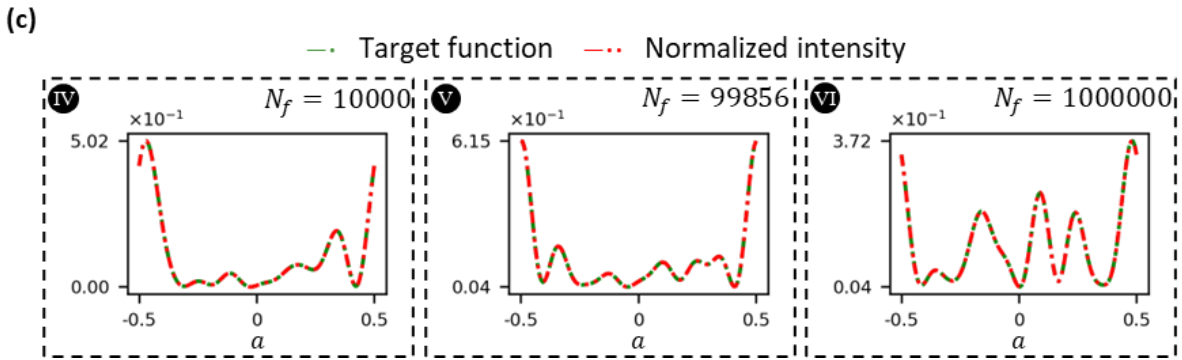

**Fig. 2: Accuracy of diffractive optical processors for increasing numbers of target nonlinear functions.** (a) Distribution of function approximation errors for five diffractive processor designs corresponding to different values of $N_f$ (number of nonlinear functions), ranging from 100 to 1,000,000. Each design uses $N \approx 1.25 \times 2 N_p N_f$ optimized and fixed phase-only features distributed over $K = 2$ diffractive surfaces, where $N_p = 9$ is the number of input pixels used to encode the argument $a$ of the functions. As $N_f$ increases, the spread and upper bound of the error

distribution also increase relatively; however, the error remains negligible across all nonlinear functions, supporting the scalability of the architecture. (b) Examples of the nonlinear functions with the minimum error from each of the three largest designs ($N_f = 10^4$, $N_f \approx 10^5$, and $N_f = 10^6$). The target functions (green curves) and their diffractive approximations (red curves) are plotted, showing perfect agreement. (c) Same as (b), but for the functions with the maximum error from each design. Even in these worst-case examples, the diffractive approximation remains indistinguishable from the target function. Also see **Supplementary Figs. S1-S3** for randomly selected function examples from the $N_f = 10^6$ set.

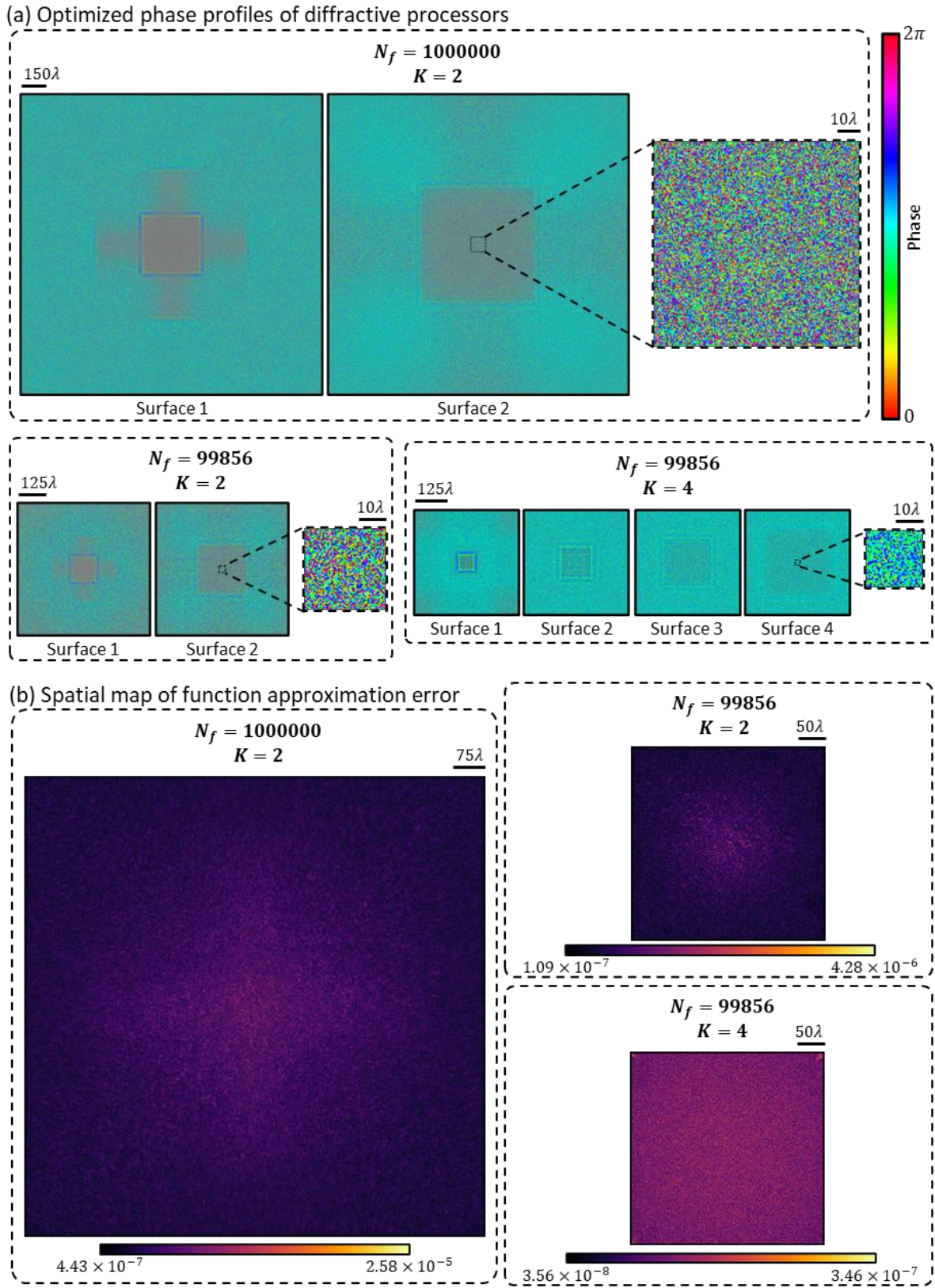

**Fig. 3: Optimized surfaces and spatial error maps of diffractive optical processors.** (a)

Optimized surface phase profiles of different diffractive processors. (b) Spatial distribution of the function approximation errors at the output planes of different diffractive processors. The approximation error of the nonlinear function corresponding to an output pixel follows the colormap. The inter-pixel gaps at the output plane are omitted here for clarity and conciseness.

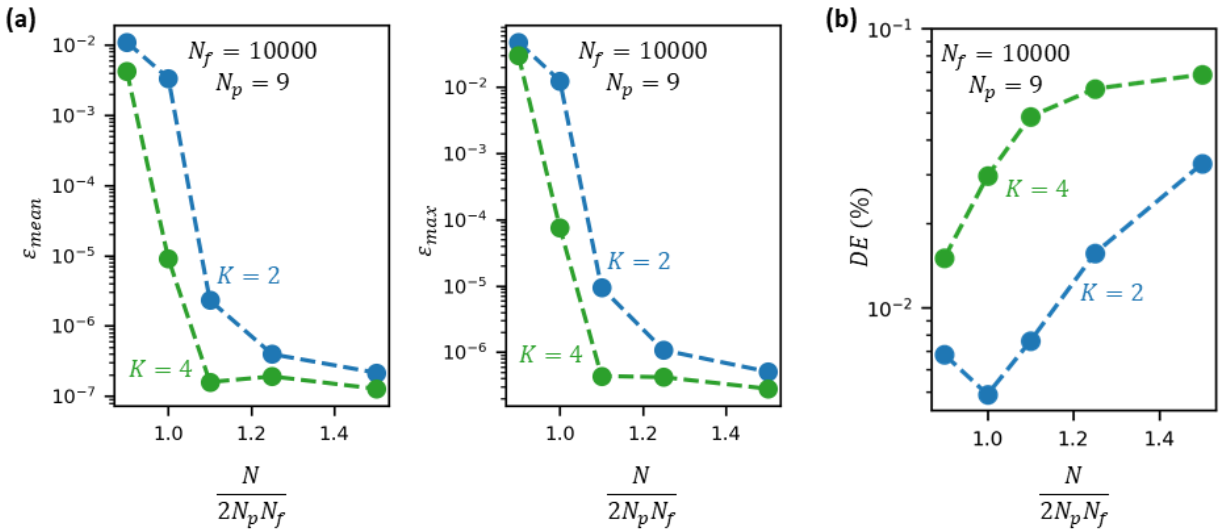

**Fig. 4: Effect of $K$ and $N$ on the nonlinear function approximation accuracy and output diffraction efficiency.** (a) Mean (left) and maximum (right) function approximation errors as a function of the total number of trainable diffractive features $N$, shown for $K = 2$ (blue) and $K = 4$ (green). Increasing the number of diffractive features reduces error, which plateaus as $N$ exceeds $2N_p N_f$, with $K = 4$ consistently achieving lower errors than $K = 2$, particularly in the low-$N$ regime. (b) Average output diffraction efficiency. Diffraction efficiency, in general, increases with $N$, and the deeper architectures ($K = 4$) yield higher diffraction efficiencies across the board.

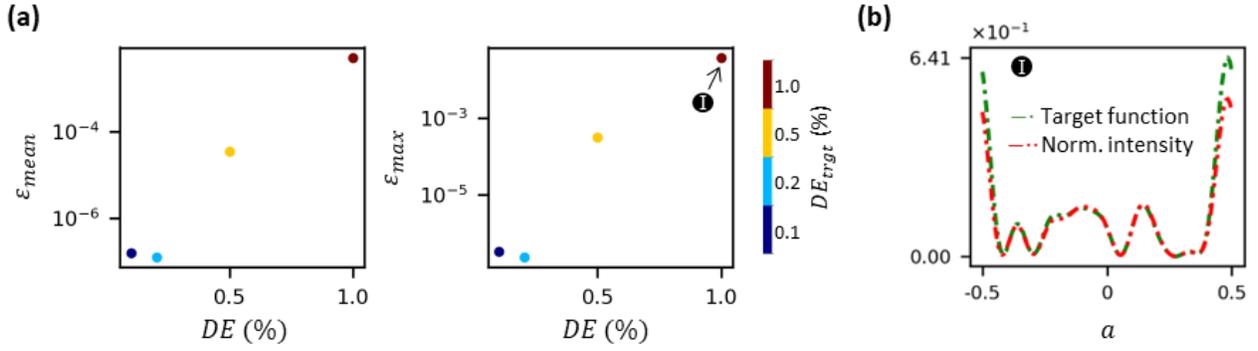

**Fig. 5: Trade-off between function approximation accuracy and diffraction efficiency.** (a) Mean and maximum function approximation errors for four diffractive processor designs, each designed with a different target diffraction efficiency $DE_{trgt}$ in the training loss function. For all these designs, $N_f = 10000$, $N \approx 1.25 \times 2N_p N_f$ and $K = 4$. Increasing $DE_{trgt}$ improves the realized diffraction efficiency but leads to a relatively higher approximation error. (b) Approximation of the function corresponding to the maximum error in the highest-efficiency design (point **I** in (a)). The diffractive approximation still closely tracks the target function, with increased deviation only near the domain edges ($a = \pm 0.5$).

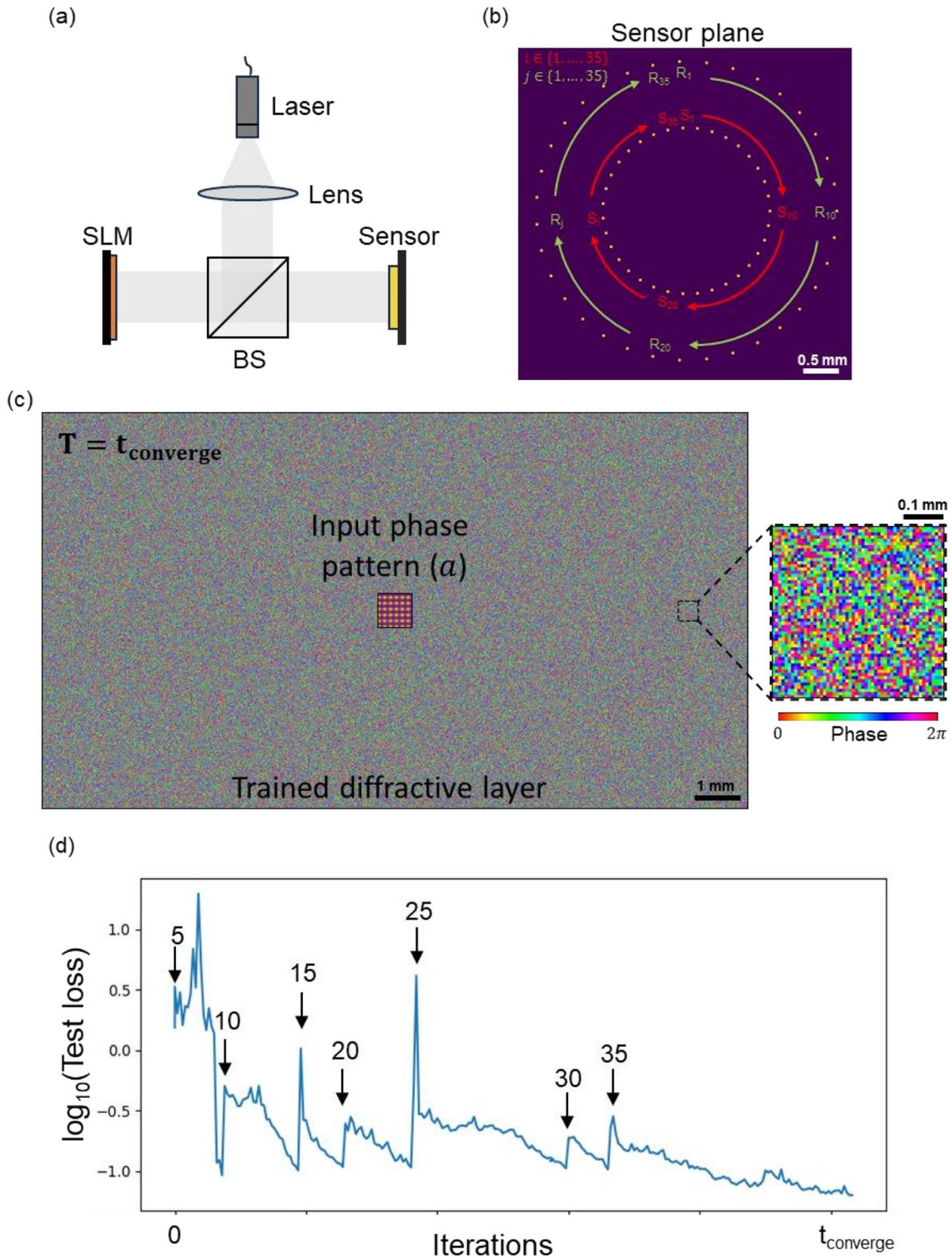

**Fig. 6: Experimental setup and *in situ* learning of a single diffractive layer to experimentally perform 35 distinct nonlinear functions, $(f_1, f_2, ..., f_{35})$.** (a) Schematic of the experimental setup

used for *in situ* learning. A collimated laser beam (520 nm) is reflected by a beamsplitter (BS) and directed onto an SLM. The input signal is encoded in the central 96×96 region of the SLM, while the surrounding pixels are trained *in situ* to realize the desired nonlinear functions. The resulting intensity distribution is captured by an image sensor for real-time feedback and post-processing. (b) Image sensor plane showing 70 predefined detection regions: 35 signal channels ($S_1$-$S_{35}$) and 35 reference channels ($R_1$-$R_{35}$). (c) Visualization of an example SLM pattern comprising the input phase pattern (for $a = 0.64$) and the learned (fixed) diffractive phase profile. (d) Test loss as a function of *in situ* training iterations. A progressive training strategy was employed, where the number of target functions was increased step by step when the loss dropped below a predefined threshold (see the **Methods** section). The downward arrows indicate the instants at which the number of functions was increased, with the associated labels (5, 10, 15, ..., 35) denoting the total number of nonlinear functions under *in situ* optimization at subsequent iterations. This progressive approach continued until all 35 nonlinear functions were jointly trained, as indicated by $t_{\text{convergence}}$.

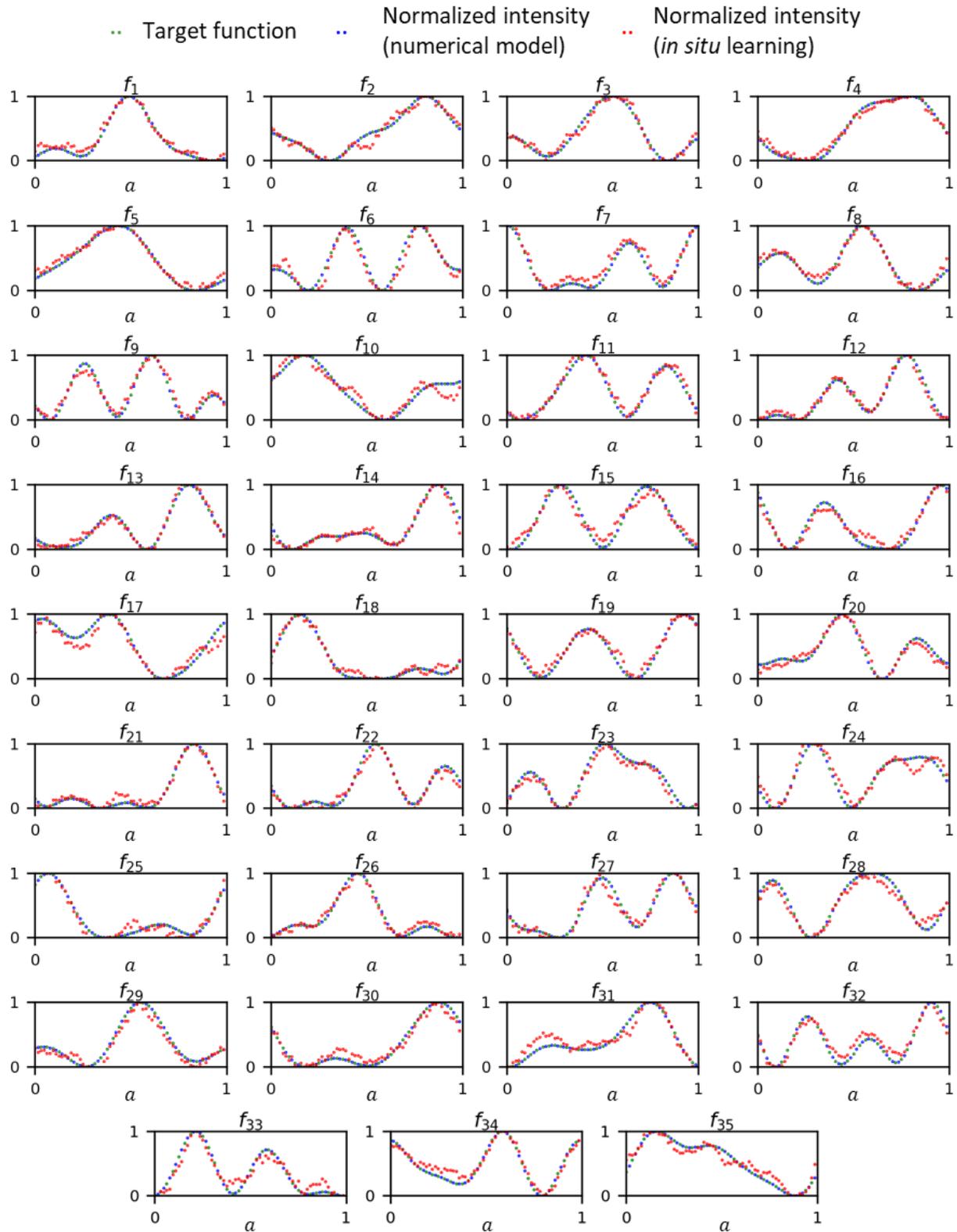

**Fig. 7: Experimentally measured outputs for 35 distinct nonlinear functions ($f_1, f_2, ..., f_{35}$).**
Once the *in situ* learning converges, the normalized output intensities follow the target nonlinear functions. In addition to the experimental *in situ* learning results, we also report numerical results based on *in silico* training, which shows near-perfect agreement with the target functions, where green and blue dots overlap. Also see **Supplementary Figs. S5** and **S6.**